\documentclass[twocolumn]{aastex62_tweeked_for_nat}

\usepackage[sort&compress]{natbib}

\usepackage[utf8]{inputenc}
\usepackage{graphicx}
\usepackage{amssymb,amsmath}
\usepackage{xcolor}
\usepackage{soul}
\usepackage{hyperref}

\begin{document}
\title{Progress in unveiling extreme particle acceleration in persistent astrophysical jets}

\author{J. Biteau} \affiliation{Institut de Physique Nucl{\'e}aire d’Orsay (IPNO), Universit{\'e} Paris-Sud, Univ.~Paris/Saclay, CNRS-IN2P3, France}

\author{E. Prandini} \affiliation{INAF -- Osservatorio Astronomico di Padova, Vicolo dell'Osservatorio 3, Padova \& INFN Sezionde di Padova, Italy}

\author{L. Costamante} \affiliation{ASI -- Unit\`a Ricerca Scientifica, Via del Politecnico snc, I-00133, Roma, Italy}

\author{M. Lemoine} \affiliation{Institut d'Astrophysique de Paris, CNRS - Sorbonne Université, 98 bis boulevard Arago, F-75014 Paris, France}

\author{P. Padovani} \affiliation{European Southern Observatory, Karl-Schwarzschild-Str. 2, D-85748 Garching bei M{\"u}nchen, Germany}

\author{E. Pueschel} \affiliation{DESY, Platanenallee 6, 15738 Zeuthen, Germany}

\author{E. Resconi} \affiliation{Technische Universit{\"a}t M{\"u}nchen, Physik-Department, James-Frank-Str. 1, D-85748 Garching bei M{\"u}nchen, Germany}

\author{F. Tavecchio} \affiliation{INAF -- Osservatorio Astronomico di Brera, via E. Bianchi 46, I-23807, Merate, Italy}

\author{A. Taylor} \affiliation{DESY, Platanenallee 6, 15738 Zeuthen, Germany}

\author{A. Zech} \affiliation{LUTH, Observatoire de Paris, PSL Research University, CNRS, Universit{\'e} de Paris, 5 Place Jules Janssen, 92190 Meudon, France}

\date{\today}

\begin{abstract}
The most powerful persistent accelerators in the Universe are jetted active galaxies. Blazars, galaxies whose jets are directed towards Earth, dominate the extragalactic $\gamma$-ray sky. Still, most of the highest-energy particle accelerators likely elude detection. These extreme blazars, whose radiated energy can peak beyond 10\,TeV, are ideal targets to study particle acceleration and radiative processes, and may provide links to cosmic rays and astrophysical neutrinos. The growing number of extreme blazars observed at TeV energies has been critical for the emergence of $\gamma$-ray cosmology, including measurements of the extragalactic background light, tight bounds on the intergalactic magnetic field, and constraints on exotic physics at energies inaccessible with human-made accelerators. Tremendous progress has been achieved over the past decade, which bodes well for the future, particularly with the deployment of the Cherenkov Telescope Array. 
\end{abstract}


\section{Introduction}
The luminosity of most galaxies is dominated by thermal emission from stars. For the remaining ${\gtrsim}\,1\%$, emission from the active galactic nucleus (AGN), which hosts a super-massive black hole ($10^6 - 10^{10}\, M_\odot$), can outshine the billions of stars in the galaxy~\cite{Padovani_2017a}. Up to ${\sim}\,10\%$ of AGN develop two-sided relativistic jets that emit non-thermally over the whole electromagnetic spectrum~\cite{Urry_1995}.  The emitting region in the jet moves away from the super-massive black hole with a relativistic bulk velocity that, for small viewing angles ($\theta < 10 - 15^\circ$), shifts the luminosity of the source to higher frequencies by a bulk Doppler factor $\delta_{\rm D} = \mathcal{O}(10)$, and amplifies the bolometric emission by $\delta_{\rm D}^4$. Such an AGN is called a blazar.

The broad-band spectral energy distribution (SED) of blazars is characterized by two distinctive humps. The first hump, peaking at infrared to X-ray wavelengths, is commonly explained as synchrotron emission from electrons accelerated in the jet. The second hump, peaking above MeV energies, is often attributed to inverse-Compton scattering, possibly of the same electrons on their own synchrotron emission (synchrotron self-Compton or SSC model, \emph{e.g.},~\cite{2010MNRAS.401.1570T}). The peak frequency of each hump, $\bar \nu$, arises from a break in the spectrum of the emitting particles, probing the comparative strength of acceleration, cooling, and escape mechanisms.

Blazars form a continuous sequence from low- to high-energy peaked objects, with an overall correlation of the two peak frequencies.  The end of the sequence is dominated by high-frequency-peaked BL~Lac objects (HBL,~\cite{pg95}), \emph{i.e.}, blazars with weak emission lines and radiatively inefficient accretion disks. 
HBLs are characterized by synchrotron emission peaking at $\bar \nu_{\rm x}\gtrsim 10^{15}$\,Hz (${\sim}\,4$\,eV), with the most \emph{extreme blazars} showing larger values by at least two orders of magnitude~\cite{1999APh....11...11G}. 

Standard models predict a $\gamma$-ray peak below 1\,TeV for HBLs, either because of the limited maximum energy of the electrons, traced by $\bar\nu_{\rm x}$, or of a reduced scattering cross-section (Klein-Nishina regime). The discovery of extreme blazars with a $\gamma$-ray peak up to 10\,TeV therefore came as a surprise. Only 3 out of 10 extreme blazars with TeV emission were detected in the GeV range a decade ago~\cite{2010MNRAS.401.1570T}. Their intrinsic properties were bound by upper limits on absorption by the extragalactic background light (EBL,~\cite{nature06}). A new window on extreme blazars was opened by the growth of $\gamma$-ray astronomy. Observational progress promoted not only developments of elaborate acceleration and radiative schemes but also the investigation of ties with cosmic magnetism, ultra-high-energy cosmic rays (UHECRs), and physics beyond the Standard Model.

\vspace{0.5cm}

\section{Extreme observational properties}

\subsection{Multi-wavelength observations of extreme blazars}

Extreme blazars come in two flavours: \emph{extreme-synchrotron} sources show a synchrotron peak energy  $h\bar \nu_{\rm x}\geq1\,$keV  ($2.4\,{\times}\,10^{17}\,$Hz); \emph{extreme-TeV} sources have a $\gamma$-ray peak energy $h\bar\nu_{\gamma}\geq1\,$TeV ($2.4\,{\times}\,10^{26}\,$Hz) (\textit{e.g.},~\cite{Costamante_2001,Costamante2018}).  
This is revealed by a hard spectrum in the soft X-ray band with photon index $\Gamma_{\rm x} < 2$, or in the $0.1-1$\,TeV band, $\Gamma_{\rm \gamma}<2$. The intrinsic spectrum is recovered from the observed spectrum by accounting for absorption, namely photoelectric absorption in intervening
gas for X-rays and $\gamma$-EBL interactions at GeV-TeV energies (see Sec.~\ref{sec:cosmo}).

In the optical band,  measurements of the flux and polarization of extreme blazars are generally hampered by the thermal emission of the giant elliptical host galaxy. These sources show radio properties similar to other HBLs, but their flux is generally very low. Indeed, a high X-ray to radio flux ratio is a very effective way to select \emph{extreme-synchrotron} sources among normal HBLs (see Sec.~\ref{catalogues}). The combination of an \emph{extreme-synchrotron} nature with a hard GeV spectrum appears to be a multi-wavelength marker favoring the selection of \emph{extreme-TeV} sources.

\begin{figure*}[ht]
\centering
\includegraphics[width=\linewidth]{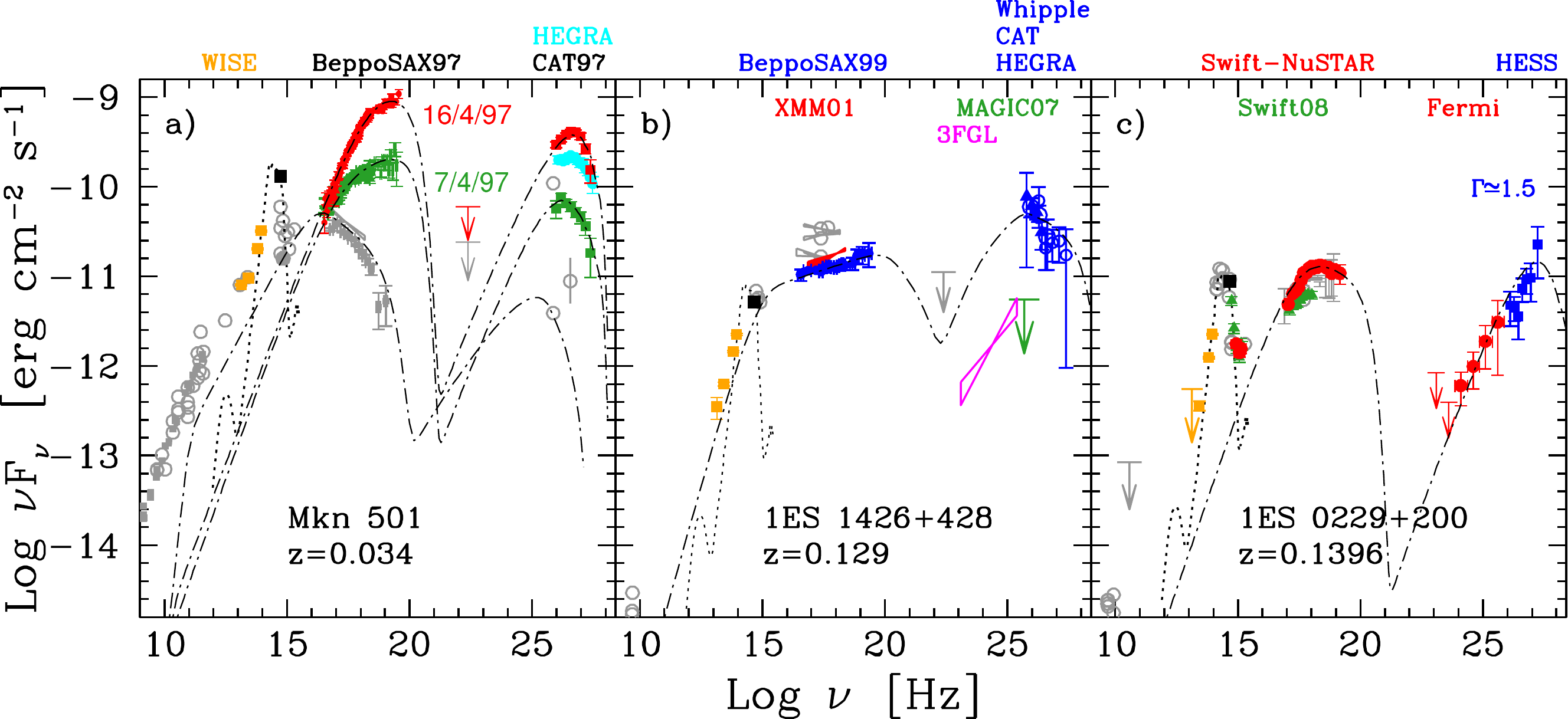}
\caption{{\bf Prototypical SEDs illustrating the three types of extreme behaviours}. Panels a) and b) display the SEDs of two \emph{extreme-synchrotron} blazars, during short-lived large flares and long-lasting quiescent states, respectively. The SED of an \emph{extreme-TeV} blazar is shown in panel c). Observed $\gamma$-ray spectra are corrected for EBL absorption~\cite{franceschini17}. Nearly-contemporaneous historical data are shown with the same colour and labelled accordingly. Grey points correspond to archival data. Error bars denote $1\,\sigma$ statistical uncertainties. Arrows denote upper limits at the 95\% confidence level (CL). Guideline SSC models are presented as dash-dotted lines. Dotted lines show the host-galaxy emission, modelled with the template spectrum of a giant elliptical galaxy~\cite{silva}, scaled to match the magnitude of the host galaxy~\cite{scarpa2000}. Data and models are from~\cite{pian98,hegra97,tavecchio01} for Mkn\,501,~\cite{ghisellini02} for 1ES\,1426+428 and~\cite{Costamante2018} for 1ES\,0229+200.} 
\label{Fig:observation}
\end{figure*}

Extreme blazars were discovered in 1996-97, with \emph{Beppo}SAX observations of flares from Mkn\,501~\cite{pian98} and 1ES\,2344+514~\cite{giommi2000}. 
Mkn\,501 reached synchrotron peak frequencies $h\bar \nu_{\rm x}>\,100\,$keV, with an increase of the luminosity by a factor of 20 in a few days and a typical ``harder-when-brighter" behaviour (Fig.~\ref{Fig:observation}, left). 
 \emph{Beppo}SAX also discovered in 1999 the first example of extreme source in low or quiescent state, with 1ES\,1426+428 displaying a hard and straight power-law spectrum over three decades in energy~\cite{Costamante_2001}. With observed $h\bar \nu_{\rm x}\gtrsim100\,$keV,  Mkn\,501 and 1ES\,1426+428 can be considered prototypical of the \emph{extreme-synchrotron} subclass.

The first \emph{extreme-TeV} blazars were discovered in 2006 with H.E.S.S.~\cite{nature06}. 
The intrinsic spectra of 1ES\,1101-232 and H\,2356+304 were unusually hard even after accounting for absorption
with the lowest possible EBL intensity, locating the $\gamma$-ray peak definitively above $1-3$\,TeV.   With the present knowledge of the EBL, eleven blazars with measured redshift can be classified as \emph{extreme-TeV} objects, as listed in Table~\ref{table:sources} (see also TeVCat~\cite{tevcat} and references therein). Three other blazars have shown \emph{extreme-TeV} properties but only temporarily, namely Mkn\,501, 1ES\,1727+502, and 1ES\,1426+428. \emph{Extreme-TeV}  properties were also found in IC\,310, a peculiar nearby AGN ($z=0.0189$) whose jet might be misaligned by $10-20^\circ$ from the line of sight~\cite{magic310}. The record for the highest $\gamma$-ray peak energy observed is held by the prototypical \emph{extreme-TeV} blazar 1ES\,0229+200 (see Fig.~\ref{Fig:observation}), with $h\bar\nu_{\gamma} \gtrsim 10\,$TeV and an intrinsic spectrum with $\Gamma_{\rm \gamma}\simeq1.5$~\cite{hess0229}. 

Most \emph{extreme-TeV} sources appear steady over years at TeV energies, although large statistical uncertainties make it difficult to rule out flux variations such as those seen in X-rays over similar timespans. The only firm evidence was found in 1ES\,1218+304, with a fast flare over a few days reaching $\sim$\,20\% of the Crab nebula flux~\cite{Acciari2010}. There are also indications of a factor of $2-3$ variations in 1ES\,0229+200  on year timescales~\cite{Aliu2014}. In both cases, the TeV spectrum remained hard, indicating no change in the \emph{extreme-TeV} character of these objects. On the contrary, during a long active phase in 2012, Mkn\,501 showed both synchrotron and $\gamma$-ray peaks shifted from normal to extreme energies, by  $\sim$\,3 and $\sim$\,1 decade in energy respectively~\cite{magic18}. The synchrotron peak of 1ES\,1727+502 similarly shifted between 2011 and 2013~\cite{2015ApJ...808..110A}.

The relationship between \emph{extreme-synchrotron} and \textit{extreme-TeV} blazars is not well established. There are 24 extreme blazars with firm redshift and published TeV spectrum (see Table~\ref{table:sources}), all of which are now detected at GeV energies by \textit{Fermi}-LAT~\cite{2019arXiv190210045T}. Twenty three of these are \emph{extreme-synchrotron} sources, at least half of which (13/23) are also \emph{extreme-TeV} blazars. On the other hand, thirteen of the fourteen known \emph{extreme-TeV} blazars are also \emph{extreme-synchrotron} in same-epoch observations or in other epochs. 
This suggests that \emph{extreme-TeV} sources may have a greater tendency to display \emph{extreme-synchrotron} behaviour at some point in time than the other way round. 

\newpage
As illustrated in Fig.~\ref{Fig:observation}, the emerging picture is that there are three distinct extreme behaviours:
\begin{enumerate}
\item becoming extreme during large flares, when both peaks shift to higher energies (Mkn\,501-like). These objects revert back to their standard HBL state; 
\item showing a steady, hard synchrotron spectrum up to $10-100\,$keV (1ES\,1426+428-like), which is not accompanied by a persistently hard TeV spectrum;  
\item showing a persistently hard $\gamma$-ray spectrum with a peak above several TeV, which remains mostly unchanged across flux variations (1ES\,0229+200-like). Their synchrotron spectrum tends to peak in the X-ray band.
\end{enumerate}

\subsection{Catalogues and population studies}\label{catalogues}

Population studies of extreme blazars are currently restricted to the \emph{extreme-synchrotron} subclass, due to the lack of an unbiased extragalactic survey at TeV energies.

The largest catalogue of extreme blazars to date is the 3HSP, which builds on the 1WHSP and the 2WHSP (see~\cite{Chang_2019} and references therein). The 1WHSP catalogue assembled a large sample of HBLs starting from the ALLWISE infrared survey and applying further restrictions on their infrared slope, as well as on their infrared/radio and infrared/X-ray flux ratios. The 2WHSP abandoned the infrared slope criterion, as it excluded sources dominated by the host galaxy. The 3HSP catalogue complemented the 2WHSP sample with \emph{Fermi}-LAT $\gamma$-ray sources with hard spectra ($\Gamma_\gamma <2$) and with objects with a high X-ray to radio flux ratio, thus selecting HBLs. The final sample includes 2,011 HBLs, with redshift information for 88\% of them (including photometric estimates), out of which 199 qualify as \emph{extreme-synchrotron} blazars. All presently known \emph{extreme-TeV} blazars are part of the 3HSP.

 \textit{Fermi}-LAT has now detected half of the ${\sim}\,340$ 3HSP sources with $\bar \nu_{\rm x}\geq10^{17}\,$Hz~\cite{2019ApJ...882L...3P}. Only a sub-sample of the studied 3HSP sources, with a higher $\bar \nu_{\rm x}$ and harder $\Gamma_\gamma$, constitute the \textit{extreme-TeV} blazar population, whose size and properties remain to be determined.

\begin{figure} [t]
\centering
\vspace{0.3cm}
\includegraphics[width=\linewidth]{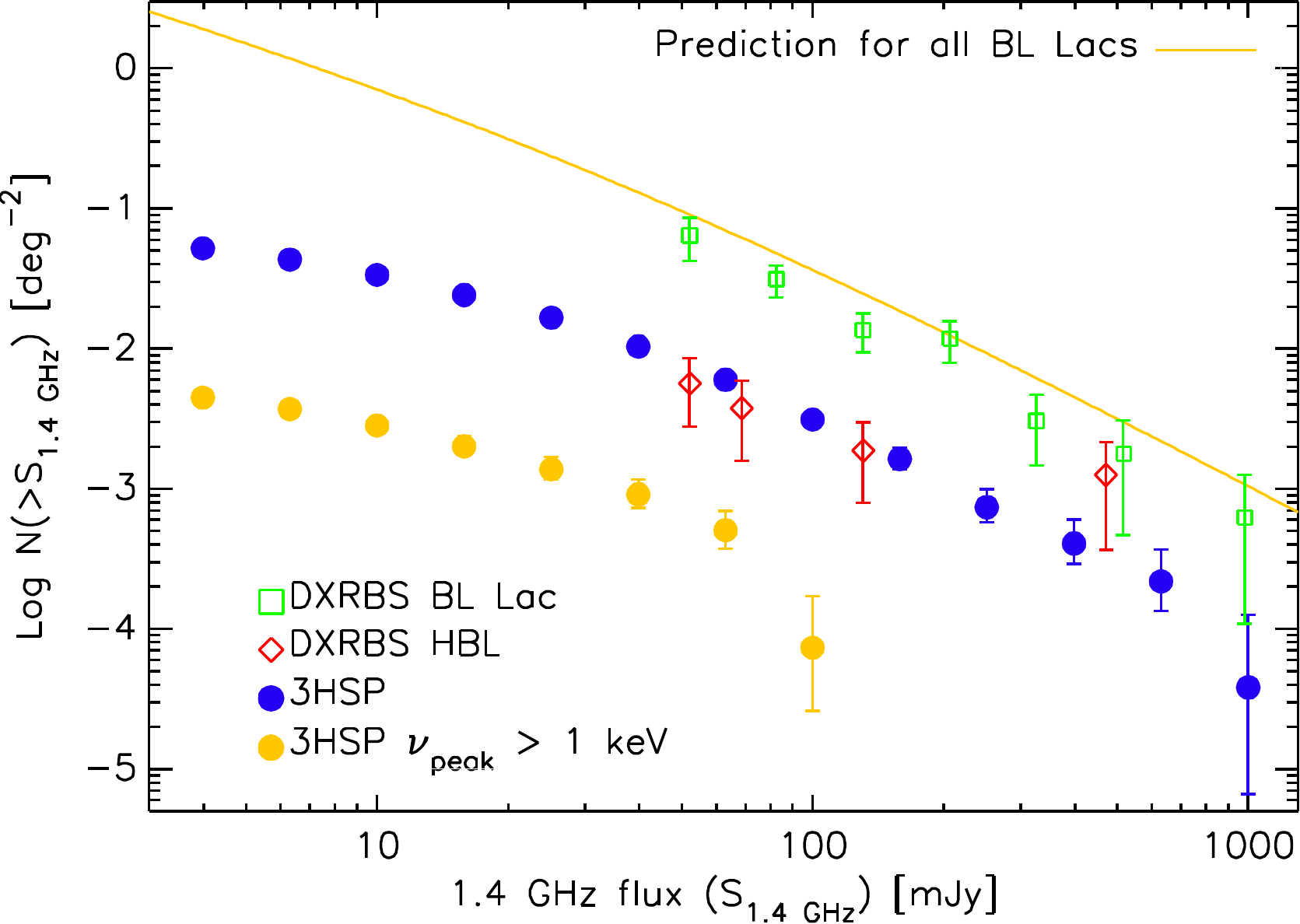}
\caption{{\bf Radio number counts of \emph{extreme-synchrotron} blazars and of different samples of BL\,Lacs}. Green squares denote BL\,Lacs from the Deep X-Ray Radio Blazar Survey (DXRBS), red trapezoids the HBLs from DXRBS, blue circles the HBLs from 3HSP, and orange circles 
the \emph{extreme-synchrotron} blazars from 3HSP. Error bars denote $1\,\sigma$ statistical uncertainties. The solid line represents the number counts predicted from the luminosity function of BL\,Lacs~\cite{Urry_1995}. Figure adapted from~\cite{Chang_2019}, courtesy of Yu-Ling Chang.}
\label{3HSP_counts}
\end{figure}

The 3HSP catalogue is flux-limited in the radio and X-ray bands for sources with $\bar \nu_{\rm x}>10^{16}$\,Hz. Radio number counts are displayed in Fig.~\ref{3HSP_counts}, together with the  number counts of other BL\,Lac samples from the Deep X-ray Radio Blazar Survey~\cite{Padovani_2007}. The slopes of the sub-sample distributions are similar, implying that the fraction of high-$\bar \nu_{\rm x}$ blazars is roughly independent of radio flux. \emph{Extreme-synchrotron} blazars represent about $10\%$ of HBLs, which in turn correspond to $10\%$ of all BL\,Lacs.

Above a radio flux of $3.5$\,mJy at 1.4\,GHz, the surface density of $\sim 4.5 \times 10^{-3}~{\rm deg}^{-2}$ inferred for extreme blazars corresponds to a total number of \emph{extreme-synchrotron} sources over the sky of $\sim$\,200. The maximum observed radio luminosity tends to be lower for extreme blazars than for regular HBLs. This is consistent with the inferred anti-correlation between power and $\bar \nu$ often denoted as the ``blazar sequence''~\cite{Ghisellini_2017}, which can also be explained by selection effects~\cite{giommi12}. The rareness of \emph{extreme-synchrotron} sources suggests the latter, as these sources, which are at the low end of the luminosity function, would otherwise be expected to represent the most common subclass along the blazar sequence.

\section{The challenge of modeling\\ extreme blazars}
\label{Sec:modeling}

The simplest one-zone SSC models generally represent well the stationary SED of most HBLs with $h\bar\nu_{\gamma}<1\,$TeV~\cite{2010MNRAS.401.1570T}, including purely \emph{extreme-synchrotron} sources.
This suggests that \emph{extreme-synchrotron} sources are the high-energy tail of the blazar population.
On the other hand, the origin of the \emph{extreme-TeV} emission is still unknown, as it presents two essential challenges:
\begin{enumerate}
\item explaining the peak of radiation at TeV energies: $h\bar\nu_{\gamma}>1\,$TeV. 
\item explaining the hard intrinsic spectrum of the component radiated at sub-TeV energies: $\Gamma_{\gamma}<2$. 
\end{enumerate}

A hard intrinsic spectrum at TeV energies, $F_\nu / \nu \propto\nu^{-\Gamma_\gamma}$ with $\Gamma_\gamma \simeq 1.5$, typically implies a hard accelerated particle spectrum, ${\rm d}N/{\rm d}\gamma\propto\gamma^{-p}$ with spectral index $p<2$, so that most of the energy is carried by the highest-energy particles, an unusual case in astrophysics. The spectral index derives from a competition between energy gains and escape or energy losses in the acceleration region. In shock acceleration, this competition leads to spectral indices $p\simeq2$ (\cite{1987PhR...154....1B};~\cite{2014ApJ...783L..21S} for relativistic shocks). Additional effects, such as non-linear backreaction of the accelerated particles, could harden the spectra to $p<2$. Both $p>2$ and $p<2$ appear to be possible in other acceleration scenarios, \textit{e.g.}, turbulent acceleration~\cite{1989ApJ...336..243S}, shear acceleration (\textit{e.g.},~\cite{2019Galax...7...78R}), or reconnection (\textit{e.g.},~\cite{2015SSRv..191..545K}), since energy gains and loss rates depend on additional parameters. Ab-initio reconnection simulations have in particular now shown hard particle spectra for a large ratio of magnetic to rest-mass energy density~\cite{2015SSRv..191..545K}.

The observations of most blazars are matched by phenomenological models, which leave the acceleration scenario unspecified and fit the environmental parameters and the accelerated particle spectrum to reproduce SEDs. These models are separated into two classes: leptonic for electrons and positrons, hadronic for protons and nuclei.

\subsection{Leptonic scenarios}

In simple one-zone SSC models, TeV radiation results from inverse-Compton scattering of electrons or pairs with comoving-frame energy $\epsilon'_e$. A peak emission beyond 1\,TeV imposes $ \delta_{\rm D}\, \epsilon'_e > 1\,{\rm TeV} $ in the Klein-Nishina regime, which implies high Lorentz factors for the electrons, $\gamma'_e=\epsilon'_e/m_e c^2 > 2{\times}10^5 \times (\delta_{\rm D}/10)^{-1} $. Both the high $\bar\nu_{\gamma}$ and hard $\Gamma_{\gamma}$ of \emph{extreme-TeV} blazars are matched by such models provided the electron distribution shows a minimum Lorentz factor with a sufficiently large value, sometimes up to $\gamma_{\rm min}\simeq 10^4-10^5$ (\textit{e.g.},~\cite{Katarzynski2006a}). These models require an unusually low radiative efficiency and conditions out of equipartition by several orders of magnitude. A sub-TeV spectrum as hard as $\Gamma_{\gamma} = 2/3$  can be obtained in an environment with a very weak magnetic field ($B\lesssim1\,$mG) where spectral softening due to synchrotron cooling is reduced. 

A different approach within the leptonic framework considers additional inverse-Compton components from external photon fields. Models with a sheath surrounding an inner jet~\cite{2005A&A...432..401G,2017MNRAS.466.3544C} struggle to reproduce a high $\bar\nu_{\gamma}$, due to the intense target photon field. However, up-scattering of photons from the broad-line region in a compact emission zone~\cite{Lefa2011} or from the cosmic microwave background (CMB) in the extended kpc-jet~\cite{Boettcher2008} could produce hard spectra up to TeV energies. These scenarios require either a sufficiently luminous broad-line region, which is not expected for extreme blazars and other BL\,Lacs, or a very energetic particle distribution over kpc scales. The latter would lead to steady TeV emission over thousands of years, in contradiction with flux variations detected in 1ES\,1218+304.

\begin{figure}[ht]
\vspace{0.2cm}
\includegraphics[width=0.95\linewidth]{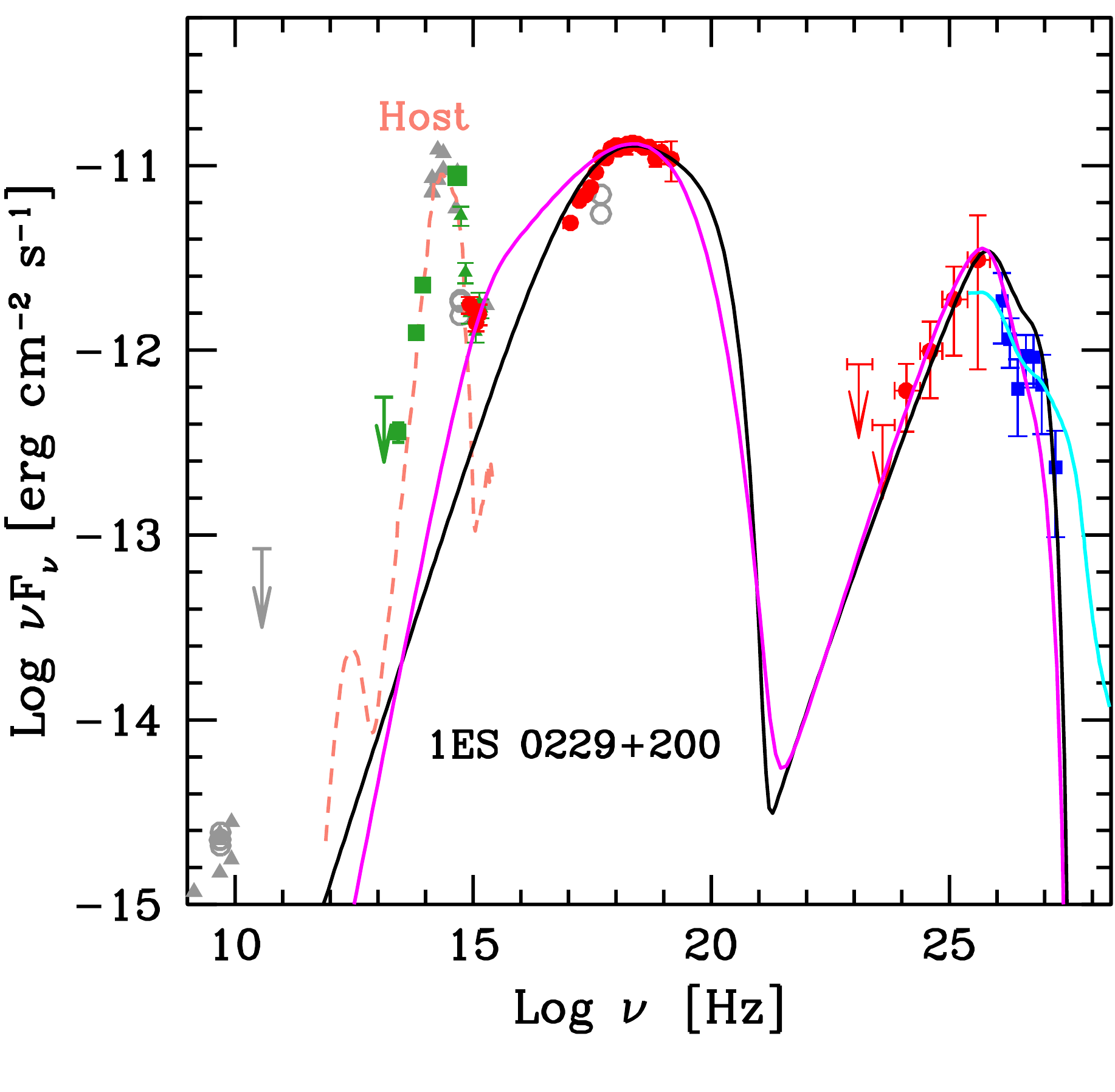}
\caption{{\bf Phenomenological models of the observed SED of the extreme blazar 1ES\,0229+200.} $\gamma$-ray data are \textit{not} corrected for EBL absorption. Error bars denote $1\,\sigma$ statistical uncertainties. Arrows denote upper limits at the 95\% CL. Representative leptonic  (SSC,~\cite{Costamante2018}) and hadronic (proton synchrotron,~\cite{Cerruti2015}) models are shown as black and magenta lines, respectively. The cyan line covering only the $\gamma$-ray range depicts the external cascade model~\cite{Murase2012}. The orange dashed line shows the emission from the host galaxy. }
\label{Fig:models}
\end{figure}

The most successful leptonic models applied to \emph{extreme-TeV} blazars thus appear to be simple SSC models that involve electron distributions with either a large minimimum Lorentz factor, $\gamma_{\rm min} \gg 1$, or hard particle spectra, $p<2$, as well as a magnetic field strength well below equipartition~\cite{Lefa2011,Tavecchio2009}. As an example, the model for 1ES\,0229+200 in Fig.~\ref{Fig:observation} and Fig.~\ref{Fig:models} assumes a moderate $\gamma_{\rm min} = 100$, a very hard particle spectrum with index $p =1.4$, a high Doppler factor $\delta_D = 50$ and a low magnetic field of 2\,mG. The electron energy density dominates over the magnetic one by a factor of $\sim 10^5$. 

The requirement of weak magnetization that limits spectral softening from radiative cooling appears to disfavour reconnection scenarios, which may otherwise produce high $\gamma_{\rm min}$ values (\textit{e.g.},~\cite{Tavecchio2009}). Stochastic acceleration could result in narrow pseudo-Maxwellian particle distributions with high peak energies (\textit{e.g.},~\cite{Tramacere2011}), although a magnetic field well below equipartition implies a slow acceleration process in this case.

\subsection{(Lepto-)hadronic models}

Typical hadronic scenarios for extreme blazars are based on the co-acceleration of protons and electrons and attribute the TeV emission to proton synchrotron radiation or to the decay of pions from $p\gamma$ interactions, possibly mixed with SSC emission from secondary pairs (\textit{e.g.},~\cite{Cerruti2015}). Proton synchrotron scenarios require ultra-high-energy particles with $\epsilon'_p > 10^{19}\,{\rm eV} \times (B'/100\,{\rm G})^{-1/2}$ in a strong comoving-frame magnetic field, $B'$~\cite{Aharonian2000}. Mixed models propose instead an energy budget dominated by the kinetic energy of protons. Such models require hard particle injection spectra, with $p$ from 1.3 to 1.7, as well as large jet energetics, typically close to the Eddington limit of the central black hole.
 
If extreme blazars accelerate very-high-energy cosmic rays, the observed TeV emission might even result from secondary radiation produced outside the source, through an electromagnetic cascade triggered by photo-pion or photo-pair production on the EBL and CMB (\textit{e.g.},~\cite{Murase2012, Essey2010, Taylor2011, Dzhatdoev2017}). 
The original version of this scenario assumed protons or nuclei with energies ${\sim}\,10^{16}$\,eV, well below the ankle of the UHECR spectrum (at a few $10^{18}\,$eV). However, nuclei at these energies are deflected by magnetic fields associated to structures of the cosmic web (voids, clusters, filaments, with magnetic fields up to 10\,nG), which broadens the beam and increases the luminosity required to reproduce the observed $\gamma$-ray spectra~\cite{Tavecchio2014}. Such models predict slower variability than that observed, although synchrotron radiation from secondary pairs in magnetic fields extending on Mpc scales could account for variability down to year timescales~\cite{Oikonomou2014}. 

As an example, Fig.~\ref{Fig:models} shows a one-zone proton synchrotron model and an external cascade model applied to 1ES\,0229+200. The proton synchrotron model requires a much larger magnetic field ($B'\simeq 20$\,G) than the leptonic model, a comparable value for $\gamma_{\rm min} \simeq 100$ and a smaller Doppler factor, $\delta_{\rm D} \simeq 30$. The particle spectra are again very hard and the considerable energy of the protons leads to a jet power at the Eddington limit. 

This illustrative proton synchrotron model predicts a peak neutrino flux of ${\sim}\,10^{-15}$\,erg\,cm$^{-2}$\,s$^{-1}$ around $10^{19}\,$eV. Such fluxes are beyond the reach of current-generation neutrino observatories. On the other hand, the recent indication of a link between a neutrino track event and the blazar TXS\,0506+056~\cite{IceCube:2018dnn, IceCube:2018cha}, which is not an extreme source, can be successfully modelled if a sub-dominant component of the  $\gamma$-ray emission arises instead from $p\gamma$ or $pp$ interactions~\cite{2019NatAs...3...88G,2019MNRAS.483L..12C,2018ApJ...864...84K,2019PhRvD..99f3008L}. These two hadronic channels are not expected to dominate the emission of \emph{extreme-TeV} blazars, but other classes of the blazar population have been proposed to contribute to the neutrino and cosmic-ray background up to the ankle (\textit{e.g.},~\cite{2014ApJ...793L..18T, 2015MNRAS.452.1877P}).

Whether AGN jets can accelerate UHECRs above the ankle remains unresolved. 
Interestingly, the high $\bar\nu_{\rm x}$ of extreme sources suggests that they are the most efficient accelerators of the AGN population. Neglecting energy losses, the ratio of the acceleration timescale, $t_{\rm acc}$, to the gyration time, $t_{\rm L}$, is constrained by the peak synchrotron frequency in the comoving frame, as $t_{\rm acc}/t_{\rm L} \lesssim 150\,{\rm MeV}/ h\bar\nu'_{\rm x}$~\cite{1983MNRAS.205..593G,2012SSRv..173..341A}. Therefore, the higher $\bar \nu'_{\rm x}$, the smaller $t_{\rm acc}/t_{\rm L}$. If this property can be extrapolated to ultra-high energies, faster acceleration points to a larger maximum energy in extreme blazars than in other AGNs~\cite{1984ARA&A..22..425H}. This seems in line with phenomenological models of the UHECR spectrum and composition observables above the ankle, which suggest particle spectra with $p<1$ escaping from the sources~\cite{Aab:2014aea}, or $p \simeq 1.5-2$ for source populations with a negative evolution~\cite{Taylor:2015rla}, as inferred for HBLs and in particular extreme blazars~\cite{Chang_2019,2014ApJ...780...73A}. 

At even higher energies, interactions with the EBL and the CMB constrain the sources of UHECRs at ${\sim}\,10^{20}\,$\,eV to lie within $100\,$Mpc, or even closer for nuclei lighter than iron and heavier than protons~\cite{1984ARA&A..22..425H,2011PhRvD..84j5007T}. Although the nearest extreme blazar, Mkn\,501, is at a luminosity distance of ${\sim}\,150$\,Mpc, misaligned counterparts viewed as radio galaxies could satisfy the local proximity constraint.
Because of a tight lower bound on the magnetic luminosity (\textit{e.g.},~\cite{2009JCAP...11..009L}, or~\cite{2018MNRAS.479L..76M} for lower-frequency-peaked radio galaxies such as Cen\,A), only luminous sources should be able to reach the highest energies, so that nearby ones may be identifiable in recent electromagnetic surveys~\cite{2019FrASS...6...23B}.

\section{Extreme blazars and tests of\\ $\gamma$-ray propagation }

\begin{figure}[ht]
\vspace{0.2cm}
\includegraphics[width=0.95\linewidth]{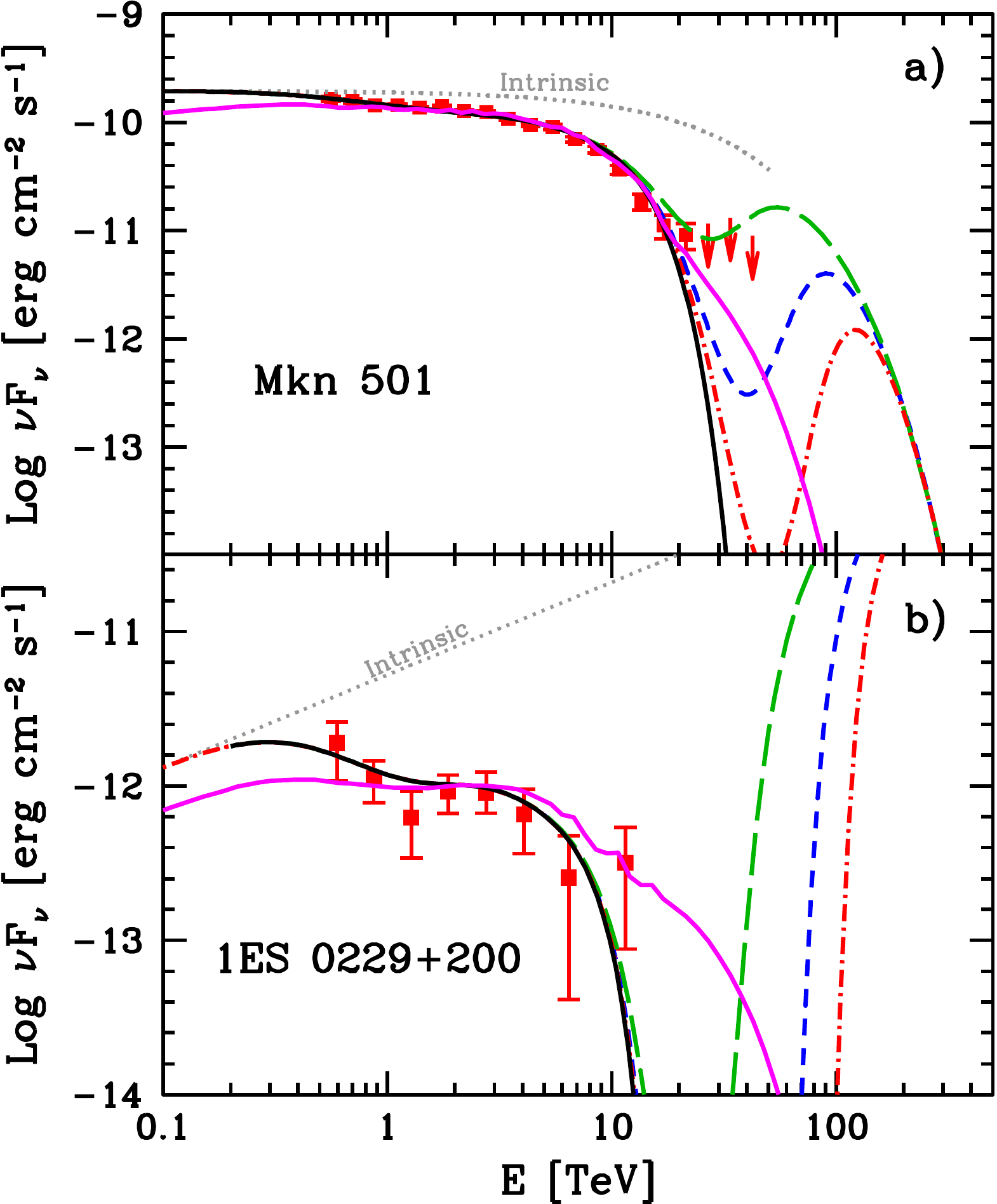}
\caption{{\bf Observed TeV spectra of two extreme blazars including different $\bm \gamma$-ray propagation models}. The steady emission of 1ES\,0229+200~\cite{hess0229} and the active phase of Mkn\,501 in 1997~\cite{hegra97} are shown in panels a) and b), respectively. $\gamma$-ray data are \textit{not} corrected for EBL absorption. Error bars denote $1\,\sigma$ statistical uncertainties. Arrows denote upper limits at the 95\% CL. The solid black line shows the expected spectra after interaction with the EBL, assuming the intrinsic emission shown by the dotted grey line. The  modifications induced by Lorentz invariance violation~\cite{TavecchioBonnoli16} are shown by the long-dashed green, dashed blue and dash-dotted red lines for $E_{\rm LIV}=3\times10^{19}$, $10^{20}$ and $2\times 10^{20}$\,GeV, respectively. The magenta solid line shows the spectrum modified by mixing with axion-like particle~\cite{Galanti19}. } 
\label{fig:fund}
\end{figure}

 \subsection{Cosmology: Extragalactic Background Light}
 \label{sec:cosmo}

The $\gamma$-ray spectra of blazars are softened by pair-production interactions with the EBL above ${\sim}100\,$GeV, in an energy- and redshift-dependent fashion shown in Fig.~\ref{fig:fund}  (\textit{e.g.},~\cite{2001ARA&A..39..249H}). The first component of EBL, peaking at optical frequencies, is now well determined thanks to both direct modeling (\textit{e.g.},~\cite{franceschini17}) and $\gamma$-ray constraints (\textit{e.g.},~\cite{2015ApJ...812...60B,desai19}). Emission from \emph{extreme-TeV} blazars detected around 10\,TeV provides constraints on the EBL intensity up to ${\sim}\,50\,\mu$m. The limited number of \emph{extreme-TeV} sources has thus left the second hump of the EBL (mid- to far-infrared) poorly constrained. 

EBL measurements based on blazars with a firm redshift rely on assumptions about the intrinsic emission prior to absorption. In the absence of full SED modelling, intrinsic spectra are assumed to be well-described by power-law or log-parabola functions with the flux decreasing at the highest energies. The lack of bright thermal components associated to a strong broad line region or a dusty torus supports a minimal absorption within the source (jet and galaxy nucleus).

A major step forward came with the H.E.S.S. spectral measurements of 1ES\,1101-232 and H\,2356+304 in 2006~\cite{nature06}. Their hard spectra provided the first strong constraints to the maximum intensity of the EBL. Since the discovery of the EBL imprint~\cite{2012Sci...338.1190A,2013A&A...550A...4H}, state-of-the-art TeV measurements have relied on observations of both steady and temporary \emph{extreme-TeV} blazars to constrain the longest wavelengths~\cite{HESSEBL, MAGICEBL}. 

\subsection{Cosmology: Intergalactic Magnetic Field}
 \label{sec:cosmoigmf}

With their hard and high-energy emission, \emph{extreme-TeV} blazars offer a unique possibility to constrain the presence of cascades in the intergalactic magnetic field (IGMF). The strength of the IGMF is not well constrained, despite its relevance (\textit{e.g.},~\cite{2013A&ARv..21...62D} for a review). It may have provided the seed field for magnetic fields in galaxies and galaxy clusters, and measurement of its strength and correlation length could distinguish between generation scenarios (in the early Universe or after the formation of the first galaxies). 

An electromagnetic cascade can develop when electron-positron pairs generated by $\gamma$-EBL interactions upscatter CMB photons to GeV energies. All the energy flux absorbed at TeV energies is reprocessed into the GeV band. The cascade develops when inverse-Compton scattering proceeds faster than plasma cooling of the pair beam, which depends on the primary $\gamma$-ray spectrum, $\Gamma_\gamma$~\cite{Vafin2018}. The deflection of electrons and positrons in the IGMF angularly broadens the cascade and introduces a time delay with respect to primary $\gamma$-rays, thereby affecting the low-energy $\gamma$-ray component. 

The first limits, which depend on the assumed source variability and duty cycle, came from $\gamma$-ray observations of 1ES\,0229+200 (\textit{e.g.},~\cite{2010Sci...328...73N}). Spectral observations in the GeV-TeV range constrain the strength of the IGMF to be larger than a few femto-Gauss, while searches for angular broadening have constrained intervals of even higher magnetic field~\cite{FermiIGMF, HESSIGMF, VERITASIGMF}. The predicted cascade component at GeV energies is particularly large for extreme blazars, due to their hard intrinsic spectra, and to the relative stability of their flux, which reduces uncertainties introduced by arrival-time delays. The strongest individual lower limits from \emph{Fermi}-LAT come from 1ES\,0229+200 and 1ES\,0347-121, while the VERITAS and H.E.S.S. exclusion regions make use of observations of 1ES\,1218+304, 1ES\,1101-232 and 1ES\,0229+200.

 \subsection{Fundamental physics}

Due to their emission at multi-TeV energies, extreme blazars have been identified as prime candidates for the search for TeV spectral anomalies, as theorized for Lorentz invariance violation~\cite{Liberati13} or mixing with axion-like particles~\cite{DeAngelis11}. 
 
Phenomenologically, Lorentz invariance violation leads to modifications of the dispersion relations of particles and photons, implying in particular the alteration of the pair production threshold above a $\gamma$-ray energy $E_{\rm mod}\simeq [(2m_ec^2)^2E_{\rm LIV}^n/(1-2^{-n})]^{1/(n+2)}$, where $n$ is the order of the correction and $E_{\rm LIV}$ is the energy scale at which Lorentz invariance is violated~\cite{2015ApJ...812...60B}. Above $E_{\rm mod}$, pair production is virtually forbidden and the Universe becomes transparent. The first quantitative limits have emerged from the very-high states of Mkn\,501~\cite{2015ApJ...812...60B,Abdalla19}. Complemented by the approach based on time-of-flight, extreme blazars constrain $E_{\rm LIV}$ to be above the Planck scale ($1.2\,{\times}\,10^{19}\,$GeV) for $n=1$~\cite{2019PhRvD..99d3015L}, indicating that effects would potentially be detectable only above $E_{\rm mod} \simeq 50$\,TeV. 
 
 Axion-like particles are generically predicted by several extensions of the Standard Model (such as string theory). They are expected to be light, neutral, and to couple to two photons, implying conversion of $\gamma$-rays to axion-like particles in a magnetic field.  The particle-photon oscillation during propagation could lead to ripples in the $\gamma$-ray spectra~\cite{2013PhRvD..88j2003A} and to a substantial reduction of the effective opacity at the highest energies~\cite{DeAngelis11}, as illustrated in Fig.~\ref{fig:fund}. For moderate redshifts, the reduction of the opacity is particularly relevant above few TeV. \emph{Extreme-TeV} blazars are therefore natural candidates for these studies~\cite{Galanti19}.
 
\section{Perspectives}
Three main research lines should guide future observations of extreme blazars: (i) population studies, featuring the characterization of a large number of extreme blazars at TeV energies; (ii) emission mechanisms, including the measurement of variability timescales in the TeV band; (iii) jet hadronic content, connecting extreme blazars and multi-messenger observations.

The current knowledge of the extreme-blazar population is based on surveys from radio wavelengths to X-rays. eROSITA will further enlarge the \emph{extreme-synchrotron} population with the first imaging all-sky survey up to 10\,keV~\cite{2012arXiv1209.3114M}. IXPE will open the X-ray polarimetric window, probing the magnetic fields and jet content of extreme blazars~\cite{2019arXiv190409313K}. At higher energies, future MeV telescopes such as AMEGO and e-ASTROGAM (\textit{e.g.},~\cite{2018JHEAp..19....1D}) have the potential to observe the high end of the \emph{extreme-synchrotron} population. 

As the next-generation $\gamma$-ray observatory, the Che-renkov Telescope Array (CTA,~\cite{2019scta.book.....C}) will play a central role in establishing a census of \emph{extreme-TeV} sources. Thanks to its sensitivity, CTA will likely discover numerous extremely hard objects, including sources that are too faint to be detected with \emph{Fermi}-LAT and current TeV instruments. The CTA survey of a quarter of the extragalactic sky, complemented in the multi-TeV range by LHAASO~\cite{2019arXiv190502773B} and SWGO~\cite{2019arXiv190208429A}, will provide for the first time an unbiased view on the \emph{extreme-TeV} population, blazars and their off-axis counterparts. This latter subclass of radio galaxies constitutes a promising field of research both for AGN population studies and for the search of the origin of UHECRs.

The physics of particle acceleration in extreme environments has benefited from the emergence of \emph{ab initio} numerical simulations that focus on the microscopic scales of acceleration, at the expense of idealizing on macroscopic scales (see~\cite{2015SSRv..191..545K,2015SSRv..191..519S,2019arXiv190103439W} for recent advances). Conversely, phenomenological modeling often integrates out the microphysics to deal with the macroscopic scales of the source. A major challenge of the coming years lies in bridging this gap, including \emph{e.g.}, a realistic description of relativistic jets and of their environment, as well as propagation in intervening magnetic and photon fields. In particular, an improved knowledge of the IGMF would benefit both the study of cosmic-ray sources and tests of external cascade scenarios. 

A productive interplay between theory, phenomenology and observations will be fostered by the broad-range measurements from $\sim$\,30\,GeV to hundreds TeV offered by CTA. The improved sensitivity with respect to current-generation instruments by a factor of  $5-20$  depending on energy  will make it possible to test emission and propagation models with unprecedented accuracy. The enhanced $\gamma$-ray sensitivity and energy resolution of CTA will also make deep observations of extreme blazars a formidable probe of fundamental physics. With the long-term monitoring program of CTA dedicated to AGN, we expect to access for the first time to time-resolved, broad $\gamma$-ray spectra of \emph{extreme-TeV} blazars. TeV observations will be complemented by X-ray monitoring with \emph{Swift}-XRT and future satellites like SVOM~\cite{2011CRPhy..12..298P}.

As very-high-energy extragalactic accelerators, extreme blazars are prime candidates to investigate multi-messenger correlations with neutrinos and cosmic rays. Significant improvements in the statistics of these astroparticles and in their composition are expected from next-generation astroparticle observatories, including IceCube-Gen2~\cite{2014arXiv1412.5106I}, KM3NeT~\cite{2016JPhG...43h4001A},  for the neutrino sector, and AugerPrime~\cite{2016arXiv160403637T} and TA$\times$4~\cite{Sagawa:2015yrf} for the cosmic-ray sector. The detection of astroparticle flux excesses associated with electromagnetic sources remains a most promising avenue to understand acceleration, radiative and escape mechanisms at energies beyond the reach of human-made accelerators.\\

\newpage
\section*{Data and code availability}
All data presented in this study are included in this published article (and its supplementary information files).




\acknowledgments

The authors declare no competing financial interests. Correspondence should be addressed to Jonathan Biteau~$<$biteau(at)in2p3.fr$>$ and Elisa Prandini~$<$elisa.prandini(at)pd.infn.it$>$. JB and EPr coordinated the work and mostly contributed to Sec.~1 and 5. LC and PP mostly contributed to Sec.~2. ML, ER, AT, and AZ mostly contributed to Sec.~3. EPu and FT mostly contributed to Sec.~4. All authors discussed the material and contributed to the writing of the manuscript. EPr has received funding from the European Union's Horizon2020 research and innovation programme under the Marie Sklodowska--Curie grant agreement no 664931. EPu acknowledges the Young Investigators Program of the Helmholtz Association. FT acknowledges contribution from the grant INAF CTA--SKA ``Probing particle acceleration and $\gamma$-ray propagation with CTA and its precursors'' and the INAF Main Stream project ``High-energy extragalactic astrophysics: toward the Cherenkov Telescope Array''.\\

This review is the result of several fruitful discussions raised during the meeting \emph{eXtreme19} (22-25 January 2019, Padova, Italy). The authors, as chairs of the scientific committee and review speakers, wish to thank all the participants to the meeting: 
C.	Arcaro, B.	Balmaverde, U.	Barres de Almeida, E.	Ben\'itez, D.	Bernard, E.	Bernardini, M.	Boettcher, S.	Boula, A.	Caccianiga, C.	Casadio, I.	Christie, A.	De Angelis, L.	Di Gesu, A.	di Matteo, I.	Donnarumma, M.	Doro, T.	Dzhatdoev, V.	Fallah Ramazani, R.	Ferrazzoli, I.	Florou, L.	Foffano, L.	Foschini, N.~I.	Fraija, A.	Franceschini, G.	Galanti, M.	Gonzalez, O.	Gueta, O.	Hervet, S.	Kerasioti, F.	Krauss, M.	Kreter, G.	La Mura, R.	Lico, R.	Lopez Coto, M.	Lucchini, M.	Mallamaci, M.	Manganaro, A.	Marinelli, M.	Mariotti, K.	Nalewajko, E.	Nokhrina, F.	Oikonomou, L.	Olivera-Nieto, S.	Paiano, V.	Paliya, D.	Paneque, Z.	Pei, C.	Perennes, S.	Rain\`{o}, P.	Romano, A.	Sharma, G.	Sigl, C.	Sinnis, P.	Soffitta, A.	Spolon, B.	Sversut Arsioli, A.	Tramacere, S.	Vercellone, V.	Vittorini, and H.	Xiao.

\begin{table*}
\centering
\begin{center}
 \begin{tabular}{l c c c c c} 
 \hline
 Source Name & redshift & X-ray spectrum  &  $\gamma$-ray spectrum  & Flux at 1\,keV & Flux at 100\,GeV \\ [0.5ex] 
& $z$ &  &   & [erg\,cm$^{-2}$\,s$^{-1}$] & [erg\,cm$^{-2}$\,s$^{-1}$] \\ [0.5ex] 
 \hline\hline
Mkn\,421 & 0.031 & SH & S & $10^{-12}-10^{-9}$ & $10^{-10}$\\
Mkn\,501 & 0.034 & SH & SH & $10^{-11}-10^{-10}$ & $10^{-10.5}$\\
1ES\,2344+514 & 0.044 & SH & S & $10^{-11.5}-10^{-10.5}$ & $10^{-11}$\\
1ES\,1959+650 & 0.048 & SH & S & $10^{-11}-10^{-10}$ & $10^{-11}$\\
TXS\,0210+515 & 0.049 & H & H & $10^{-11}$ & $10^{-12}$ \\
1ES\,2037+521 & 0.053 & H & H?& $10^{-11.5}$ & $10^{-12}$ \\
1ES\,1727+502 & 0.055 & HS & SH & $10^{-11}$ & $10^{-11.5}$\\
PGC 2402248 & 0.065 & H & S?  & $10^{-11.5}$ & $10^{-12}$\\
PKS\,0548-322 & 0.069 & H & H & $10^{-11}$ & $10^{-12}$ \\
RGB\,J0152+017 & 0.08 & H & S & $10^{-11.5}$ & $10^{-12}$\\
1ES\,1741+196 & 0.084 & H & H & $10^{-11.5}$ & $10^{-11.5}$\\
SHBL\,J001355.9-185406 & 0.095 & H & S & $10^{-11}$ & $10^{-12}$ \\
1ES\,1312-423 & 0.105 & H & S & $10^{-11}$ & $10^{-12}$ \\
RGB\,J0710+591 & 0.125 & H & H & $10^{-11}$&  $10^{-12}$\\
1ES\,1426+428 & 0.129 & H & SH & $10^{-11}$ & $10^{-11.5}$\\
RX\,J1136.5+6737 & 0.1342 & H & S & $10^{-11.5}$ & $10^{-12}$\\
1ES\,0229+200 & 0.1396 & H & H & $10^{-11.5}$ &  $10^{-12}$\\
1ES\,1440+122 & 0.163 & H & S & $10^{-11}$ &  $10^{-12}$\\
H\,2356-309 & 0.165 & H & H & $10^{-11}$ & $10^{-12}$\\
1ES\,1218+304 & 0.182 & HS & H & $10^{-11}$& $10^{-11}$\\
1ES\,1101-232 & 0.186 & HS & H & $10^{-11}$ &$10^{-11.5}$ \\
1ES\,0347-121 & 0.188 & H & H &  $10^{-11}$ & $10^{-12}$\\
RBS\,0723 & 0.198 & H & S & $10^{-11.5}$& $10^{-12}$\\
1ES\,0414+009 & 0.287 & S & H & $10^{-11} $ & $10^{-11.5} $\\
\hline
\end{tabular}
\caption{{\bf Salient properties of the 24 extreme blazars with firm redshift that are presently detected at TeV energies.} The X-ray and $\gamma$-ray spectra are labelled as: H\,= hard spectrum (extreme behaviour); S\,= soft spectrum; HS\,= mostly hard spectrum, sometimes soft; SH\,= mostly soft spectrum, becoming hard when flaring. As-yet uncertain classifications are annotated with a question mark. Fluxes at 1\,keV and 100\,GeV are rounded to half a dex.}
\label{table:sources}
\end{center}
\end{table*}

\newpage
\bibliography{extreme}

\end{document}